\documentclass[twocolumn,secnumarabic,amssymb, nobibnotes, aps, prb,groupedaddress,superscriptaddress]{revtex4-2}
\usepackage{amsmath,graphicx,latexsym,times,color}
\usepackage{setspace}
\usepackage[hidelinks]{hyperref}
\usepackage{array}
\usepackage{textcomp}
\usepackage{titlesec}
\usepackage{physics}
\usepackage{gensymb}
\usepackage{nccmath}
\usepackage{empheq} 

\definecolor{blue-violet}{rgb}{0.54, 0.17, 0.89}\newcommand{\V}[1]{\ensuremath{\mathbf{#1}}} 

\let\oldtimes\times  
\renewcommand\times{{\oldtimes}}

\begin{document}
	
	\title{Prediction of stable nanoscale skyrmions in monolayer Fe$_5$GeTe$_2$}

 
	\author{Dongzhe Li}
	\email[Corresponding author: ]{dongzhe.li@cemes.fr}
	\affiliation{CEMES, Universit\'e de Toulouse, CNRS, 29 rue Jeanne Marvig, F-31055 Toulouse, France}
	
	\author{Soumyajyoti Haldar}
	\affiliation{Institute of Theoretical Physics and Astrophysics, University of Kiel, Leibnizstrasse 15, 24098 Kiel, Germany}

    \author{Leo Kollwitz}
    \affiliation{Institute of Theoretical Physics and Astrophysics, University of Kiel, Leibnizstrasse 15, 24098 Kiel, Germany}

\author{Hendrik Schrautzer}
    \affiliation{Institute of Theoretical Physics and Astrophysics, University of Kiel, Leibnizstrasse 15, 24098 Kiel, Germany}
    \affiliation{Science Institute and Faculty of Physical Sciences, University of Iceland, VR-III, 107 Reykjavík, Iceland}

    \author{Moritz A. Goerzen}
    \affiliation{Institute of Theoretical Physics and Astrophysics, University of Kiel, Leibnizstrasse 15, 24098 Kiel, Germany}
 
	\author{Stefan Heinze}
	\affiliation{Institute of Theoretical Physics and Astrophysics, University of Kiel, Leibnizstrasse 15, 24098 Kiel, Germany}
	\affiliation{Kiel Nano, Surface, and Interface Science (KiNSIS), University of Kiel, 24118 Kiel, Germany}
	
	\date{\today}
	
	\begin{abstract}

Using first-principles calculations and atomistic spin simulations, we predict stable isolated skyrmions with a diameter below 10 nm in a monolayer of the two-dimensional van der Waals ferromagnet Fe$_5$GeTe$_2$, a material of significant experimental interest. A very large Dzyaloshinskii-Moriya interaction (DMI) is observed due to the intrinsic broken inversion symmetry and strong spin-orbit coupling for monolayer Fe$_5$GeTe$_2$. We show that the nearest-neighbor approximation, often used in literature, fails to describe the DMI. The strong DMI together with moderate in-plane magnetocrystalline anisotropy energy allows to stabilize nanoscale skyrmions in out-of-plane magnetic fields above $\approx 2$~T. The energy barriers of skyrmions in monolayer Fe$_5$GeTe$_2$ are comparable to those of state-of-the-art transition-metal ultra-thin films. We further predict that these nanoscale skyrmions can be stable for hours at temperatures up to 20 K.
	\end{abstract}
	
	\maketitle
	
Magnetic skyrmions -- localized, stable spin textures with intriguing topological and dynamical properties -- have emerged as a promising avenue to realize next-generation spintronics devices \cite{fert2017magnetic,everschor2018perspective,Luo2018,gobel2021beyond,li2021magnetic,Psaroudaki2021}. During the last ten years, the main focus of the community so far has been on skyrmions in bulk systems \cite{Muhlbauer2009skyrmion,yu2011near,YangPRL_2012} and interface-based systems of ultrathin films \cite{heinze2011spontaneous,Romming2013,dupe2014tailoring,boulle2016room,meyer2019isolated}, and multilayers \cite{moreau2016additive,soumyanarayanan_2017,raju2019evolution}. Recently, magnetic skyrmions were discovered in atomically thin two-dimensional (2D) van der Waals (vdW) materials, providing an ideal playground to push skyrmion technology to the single-layer limit \cite{han2019topological,ding2019observation}. Stabilizing skyrmions in 2D magnets can avoid pinning by defects due to high-quality vdW interfaces and the possibility of easy control of magnetism via external stimuli. 

The Dzyaloshinskii-Moriya interaction (DMI), which prefers a canting of the spins of adjacent magnetic atoms, is often recognized as the key ingredient in forming magnetic skyrmions. The DMI originates from spin-orbit coupling (SOC) and relies on broken inversion symmetry. However, most 2D magnets exhibit inversion symmetry, therefore, the DMI is suppressed. Several strategies have been proposed to achieve DMI by breaking inversion symmetry. These include the family of Janus vdW magnets \cite{Liang2020,Changsong2020,du2022spontaneous}, electric field \cite{Liu2018}, and 2D vdW heterostructures \cite{wu2020neel,Park2021,wu2021van}. In particular, the Fe$_n$GeTe$_2$ family ($n = 3, 4, 5$) with high Curie temperature ($T_c$) near room temperature has been proposed as a promising candidate for magnetic skyrmions. Néel-type magnetic skyrmions are reported in Fe$_3$GeTe$_2$ heterostructures by experiments \cite{wu2020neel,yang2020creation,Park2021,wu2021van} and explained 
by \textit{ab initio} theory \cite{Dongzhe2022_fgt,Dongzhe_prb2023} in terms of the emergence of DMI at the interface. All-electrical skyrmion detection has also recently been proposed in tunnel junctions based on Fe$_3$GeTe$_2$ \cite{li2023proposal}. More recently, several experimental groups reported the observation of topological spin structures (i.e., skyrmions or merons) and magnetic bubbles in 2D vdW Fe$_5$GeTe$_2$ \cite{Gao_2022,schmitt2022skyrmionic,lv2022controllable,Fujita2022,casas2023coexistence,Michael2023}. Additionally, Fe$_5$GeTe$_2$ exhibits high $T_c$ above room temperature \cite{may2019ferromagnetism,Zhang2020}, which makes it promising for spintronics device applications. However, the quantification of individual skyrmions' stability and lifetime in Fe$_5$GeTe$_2$, crucial for device applications, has been reported neither in experiments nor in theory.

\begin{figure}[t]
	\centering
	\includegraphics[width=0.9\linewidth]{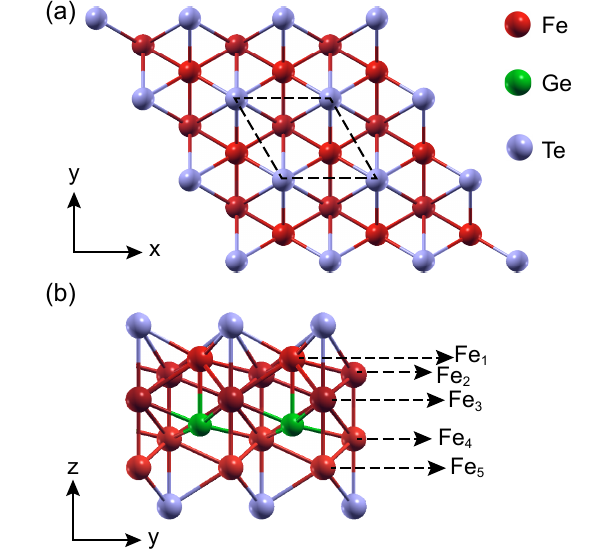}\\
	\caption{\label{F5GT_structure} (a) Top and (b) side views of the atomic structure of the Fe$_5$GeTe$_2$ monolayer. The black dashed lines draw up the 2D primitive cell. }
\end{figure}

In this Letter, we predict the formation of nanoscale skyrmions in the atom-thick vdW Fe$_5$GeTe$_2$ monolayer based on first-principles calculations and atomistic spin simulations. The diameters of these skyrmions are below 10 nm, which is technologically desirable for improving the controllability and integrability of skyrmion-based functional devices. However, such small skyrmions have not yet been observed in 2D vdW magnets. The origin of these nanoscale skyrmions is attributed to strong DMI together with moderate in-plane magnetocrystalline anisotropy energy (MAE) and weak exchange frustration. Furthermore, the calculated energy barriers of skyrmion collapse for Fe$_5$GeTe$_2$ monolayer are $\sim$80 meV at a moderate magnetic field of about 2~T. This substantial energy barrier is comparable to that of ultrathin films~\cite{Malottki2017,Haldar2018,paul2020role}, which serve as prototype systems for hosting nanoscale skyrmions. Finally, we also calculate explicitly skyrmion lifetime as a function of magnetic field and temperature. Our results demonstrate that Fe$_5$GeTe$_2$ monolayer is an excellent candidate to experimentally observe nanoscale skyrmions in a 2D vdW magnet.
	
\begin{figure}[t]
	\centering
	\includegraphics[width=1.0\linewidth]{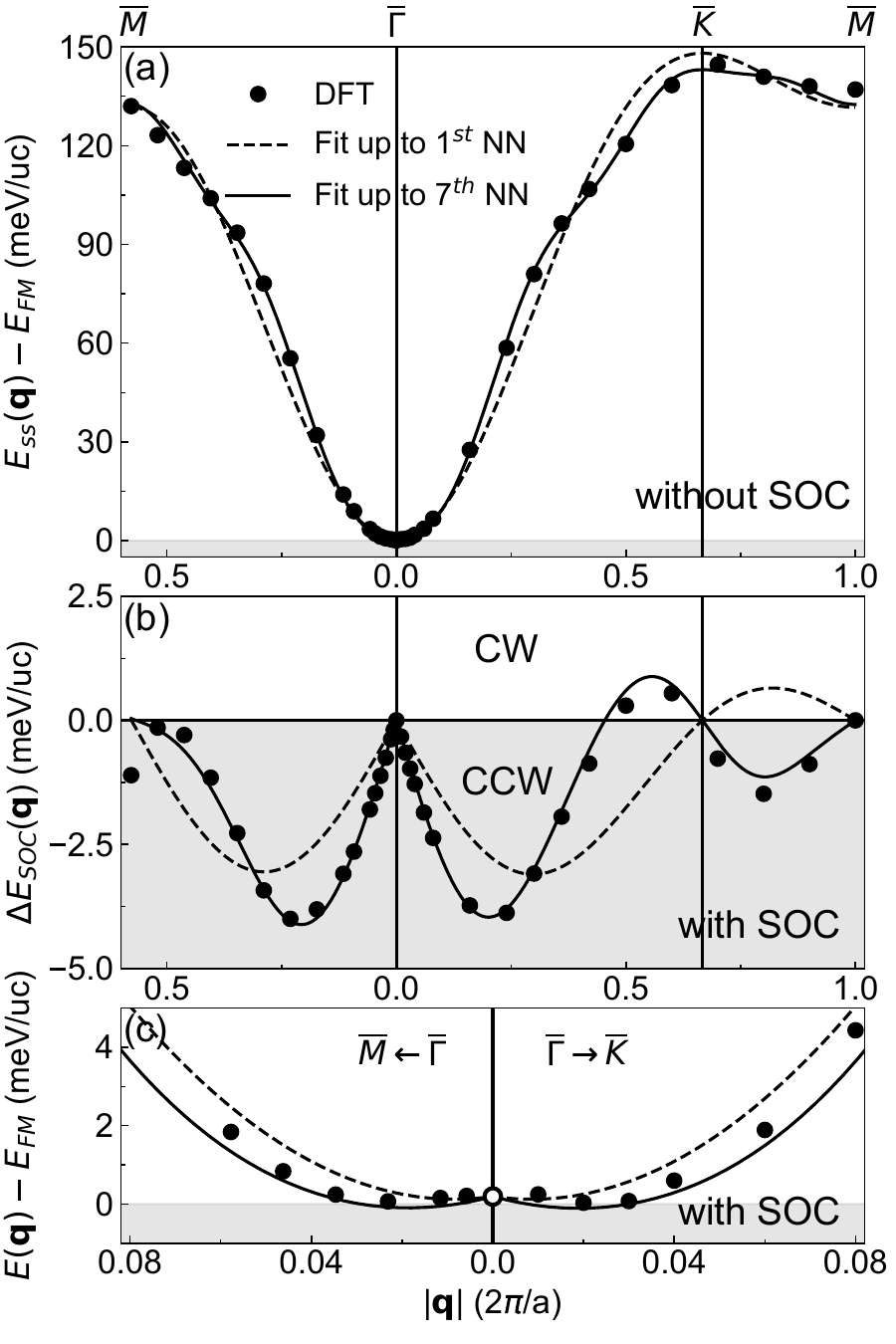}
	\caption{\label{spin_spiral} (a) Energy dispersion of flat spin spirals ($E_{\rm SS}$) for Fe$_5$GeTe$_2$ along the high symmetry path $\overline{\text{M}}$-$\overline{\Gamma}$-$\overline{\text{K}}$-$\overline{\text{M}}$ without SOC. The symbols represent the DFT calculations in scalar-relativistic approximation, while the dashed and solid lines are the ﬁts to the Heisenberg exchange interaction up to the first nearest neighbor (NN) and up to the seventh NN, respectively. (b) Energy contribution of cycloidal spin spirals due to SOC ($\Delta E_{\text{SOC}}$), also known as DMI contribution. All energies are measured with respect to the FM state ($E_{\rm FM}$) at the $\overline{\Gamma}$ point. Note that positive and negative energies represent a preference for clockwise (CW) and counterclockwise (CCW) spin configurations. (c) Zoom around the FM state ($\overline{\Gamma}$ point), including the Heisenberg exchange, the DMI, and the MAE, i.e.~$E(\V{q})=E_{\rm SS}(\V{q})+\Delta E_{\text{SOC}}(\V{q})+K/2$. The DMI leads to a CCW rotational sense, and the MAE is responsible for the constant energy shift ($K/2$) of the spin spirals with respect to the FM state.}
\end{figure}
	
\begin{table*}[htbp]
	\centering
	\scalebox{0.85}{
		\begin{tabular}{cccccccccccccccccc}
			\hline\hline
			\multicolumn{1}{c}{ } & \multicolumn{1}{c} {~~~$J_{1}/J_{\text{eff}}$~~~} & \multicolumn{1}{c} {~~~$J_{2}$~~~} & \multicolumn{1}{c} {~~~$J_{3}$~~~} & \multicolumn{1}{c} {~~~$J_{4}$~~~} & \multicolumn{1}{c} {~~~$J_{5}$~~~} & \multicolumn{1}{c} {~~~$J_{6}$~~~} & \multicolumn{1}{c} {~~~$J_{7}$~~~} & \multicolumn{1}{c} {~~~$D_{1}/D_{\text{eff}}$~~~} & \multicolumn{1}{c} {~~~$D_{2}$~~~} & \multicolumn{1}{c} {~~~$D_{3}$~~~} & \multicolumn{1}{c} {~~~$D_{4}$~~~} & \multicolumn{1}{c} {~~~$D_{5}$~~~} & \multicolumn{1}{c} {~~~$D_{6}$~~~} & \multicolumn{1}{c} {~~~$D_{7}$~~~} & \multicolumn{1}{c} {~~~$K$~~~} & \multicolumn{1}{c} {~~~$m_{s}$~~~}\\ 
			~Full model~  & 14.803  & 0.898 & 0.466 & 0.916 & 0.231 & $-$0.443 & $-$0.599 &  $-$0.659  & $-$0.523 & $-$0.083 & 0.020 & $-$0.008 & $-$0.021 & $-$0.008 & 0.380 & 7.306 \\
			~Effective model~  & 16.457  & -- & -- & -- & -- & -- & -- & $-$0.880 & --  & -- & -- & -- & -- & -- & 0.380 & 7.306 \\
			\hline
			\end{tabular}}
		\caption{Full model: Shell-resolved Heisenberg exchange constants ($J_i$) and DMI constants ($D_i$) obtained by fitting the energy contribution to spin spirals without and with SOC, and spin moments ($m_s$) from DFT calculations as presented in Fig.~\ref{spin_spiral} for the Fe$_5$GeTe$_2$ monolayer. Effective model: Parameters of the effective NN exchange and DMI model obtained by fitting the DFT results. A positive (negative) sign represents FM (AFM) coupling. A positive (negative) sign of $D_i$ denotes a preference for CW (CCW) rotating cycloidal spin spirals. The Fe$_5$GeTe$_2$ monolayer favors in-plane MAE (i.e., $K>0$). Note that all parameters are treated as a collective 2D spin model in a hexagonal symmetry (i.e., five Fe atoms in the supercell are treated as a whole). The magnetic moments are given in $\mu_{\text{B}}$/unit cell (uc), and other parameters 
  in meV/uc. 
  } \label{table_dmi}
		\end{table*}

Our ﬁrst-principles calculations were performed using the \textsc{Fleur} code \cite{fleurv26} based on the full-potential linearized augmented plane wave method (see Supplemental Material for computational details~\cite{supplmat}). We have calculated the energy dispersions 
$E(\V{q})$ of flat spin spiral states \cite{Kurz2004,Heide2009}
for the Fe$_5$GeTe$_2$ monolayer. A magnetic moment $\V{m}_{i}$ at atom position $\V{R}_i$ for a flat spin spiral is given by $\V{m}_i = M[\cos(\V{q} \cdot \V{R}_i), \sin(\V{q} \cdot \V{R}_i), 0]$, with $M$ denoting the size of the magnetic moment.

To determine the interactions of magnetic moments for the Fe$_5$GeTe$_2$ monolayer, we adopt the following atomistic spin Hamiltonian, which is fitted from spin spiral calculations without and with SOC
\begin{equation}\label{model}
\begin{split}
H & =-\sum_{ij}J_{ij}(\V{m}_i \cdot \V{m}_j)-\sum_{ij}\V{D}_{ij} \cdot(\V{m}_i \times \V{m}_j) \\
& +\sum_i K (m_i^z)^2 - \sum_i M(\V{m}_i \cdot B)
\end{split}
\end{equation}
where $\V{m}_i$ and $\V{m}_j$ are normalized magnetic moments at position $\V{R}_i$ and $\V{R}_j$ respectively. 
The four magnetic interaction terms correspond to the Heisenberg isotropic exchange, the DMI, the MAE, and the external magnetic field, and they are characterized by the interaction constants $J_{ij}$, $\V{D}_{ij}$, and 
$K$, and $B$, respectively. Note that our spin model is adapted to a collective 2D model by treating five Fe layers of Fe$_5$GeTe$_2$ as a whole system, similar to a monolayer system. All magnetic interaction parameters are measured in meV/unit cell (uc).


We consider a Fe$_5$GeTe$_2$ monolayer where one Fe$_1$ is situated above Ge as shown in Fig.~\ref{F5GT_structure}(a), the so-called UUU configuration. The calculated lattice constant is about 3.96~\AA, which agrees well with previous results \cite{YuguiYao2021}. We are aware that there are two possible configurations for monolayer Fe$_5$GeTe$_2$, namely UUU and UDU configurations \cite{ershadrad2022unusual,ghosh2023unraveling}. In this work, we focus on the UUU configuration since skyrmionic spin structures are experimentally observed in this configuration \cite{Gao_2022,schmitt2022skyrmionic}. For both spin channels, Fe$_5$GeTe$_2$ exhibits a metallic property (See Fig.~S1 in Supplemental Material~\cite{supplmat}). The calculated spin moments are $-0.21\mu_{\text{B}}$, $2.21\mu_{\text{B}}$, $1.70\mu_{\text{B}}$, $1.28\mu_{\text{B}}$, and $2.41\mu_{\text{B}}$ for the Fe$_1$, Fe$_2$, Fe$_3$, Fe$_4$, and Fe$_5$ atoms, respectively. Note that the calculated spin moments are, in general, in good agreement with previous \textit{ab initio} results, as summarized in detail in Table S2 in Supplemental Material~\cite{supplmat}. It is important to note that the spin moment on Fe$_1$ is significantly quenched, which is also in good agreement with dynamical mean-field (DMFT) results \cite{ghosh2023unraveling}.


We focus first on spin spiral calculations without SOC along the 
$\overline{\text{M}}$-$\overline{\Gamma}$-$\overline{\text{K}}$-$\overline{\text{M}}$ high-symmetry direction
(Fig.~\ref{spin_spiral}(a)). 
From the fitted parameters shown in Table \ref{table_dmi}, we find the Heisenberg exchange 
to be largely dominated by the 
nearest neighbor contribution. This is also clearly seen from the fitted curve using only the 
nearest neighbor (NN) term (see dotted line in 
Fig.~\ref{spin_spiral}(a)). On the other hand, exchange
constants up to seventh neighbors (Table \ref{table_dmi}) are necessary to fit the DFT results accurately (solid line in Fig.~\ref{spin_spiral}(a)).
Without SOC, the energy dispersion shows a minimum at the $\overline{\Gamma}$ point, which 
represents the ferromagnetic (FM) state.
The energy difference between the $\overline{\Gamma}$ 
and the $\overline{\text{M}}$ point (row-wise antiferromagnetic (AFM) state) is found to be about 137~meV for five Fe atoms. This energy difference is much smaller than 
that in the Fe$_3$GTe$_2$ or Fe$_4$GeTe$_2$ monolayers (see Fig.~S2 in Supplemental Material~\cite{supplmat}), leading to much smaller 
$J_1$ values (see Table I in Supplemental Material~\cite{supplmat}). We also note that the spin moment variation is mainly from Fe$_1$ and Fe$_4$ atoms
(Fig.~S3a in Supplemental Material ~\cite{supplmat}). Additionally, we have also carefully checked the effect of variation of spin moments for our spin model by calculating conical spin spirals (see Section II, Fig.~S3b, and Fig. S4 in Supplemental Material~\cite{supplmat}.)

When SOC is taken into account, the DMI arises due to broken inversion symmetry. The Fe$_5$GeTe$_2$ monolayer favors cycloidal spin spirals with a counter-clockwise (CCW) rotational sense, as seen from the calculated energy contribution to the dispersion due to SOC, $\Delta E_{\text{SOC}}(\V{q})$
(Fig.~\ref{spin_spiral}(b)). If we apply the NN approximation, we obtain the effective DMI constant $D_{\text{eff}} = -0.88$~meV, and the corresponding micromagnetic DMI is given by
$D=\frac{3\sqrt{2}D_{\text{eff}}}{N_{\text{F}}a^2}$ ($a$ and $N_{\text{F}}$ 
are the lattice constant and the number of ferromagnetic layers). For the Fe$_5$GeTe$_2$ monolayer, the value is approximately 0.76 mJ/m$^2$, which is in reasonable agreement with the one calculated by the supercell approach ($\sim$0.48$-$0.67 mJ/m$^{2}$) \cite{Gao_2022} based on the NN approximation. 
However, as clearly seen in Fig.~\ref{spin_spiral}(b), the dashed lines with an effective NN DMI fail to capture our DFT results beyond the regime of small $|\mathbf{q}|$.
The inclusion of interactions up to the seventh NN is needed to reproduce the DFT data accurately, and this is also reflected in the fitted parameters presented in Table \ref{table_dmi}. The second NN interaction is on the same order of magnitude compared to the first NN one.

Additionally, SOC also introduces MAE. We find the easy magnetization axis of 
the Fe$_5$GeTe$_2$ monolayer to be in-plane with 
a MAE of 0.38 meV/uc. As observed in Fig.~\ref{spin_spiral}(c), a spin spiral energy minimum of $-$0.1 meV/uc compared to the FM state occurs in the 
$\overline{\Gamma\text{K}}$ direction, which corresponds to a spin spiral period of $\lambda = 2\pi/\V{|q|} = 19.9$ nm.

Including all interactions, i.e.~Heisenberg exchange, DMI, and MAE, results in the spin spiral energy dispersion shown in Fig.~\ref{spin_spiral}(c). The energy contribution from the MAE leads to an energy offset of $K/2$ for spin spirals with a long period, i.e.~small value of
$|\mathbf{q}|$,
with respect to the FM state, as can be seen in a zoom of $E(\V{q})$ around 
$\overline{\Gamma}$. Interestingly, the ground state within the NN approximation is the FM state, while it is the spin spiral state when we include interactions up to the seventh NN. It is worth emphasizing that the spin spiral curve becomes extremely flat near the $\overline{\Gamma}$ point. It has been 
demonstrated that a flat energy dispersion around 
$\overline{\Gamma}$ is beneficial for stabilizing nanoscale skyrmions in ultrathin films \cite{meyer2019isolated}. To check the validity 
of our 2D collective spin model, we performed Monte-Carlo simulations (see section III in Supplemental
Material~\cite{supplmat}) with magnetic interaction parameters from the full model presented in Table \ref{table_dmi}. We obtain the Curie temperature ($T_c$) of about 451 \text{K} (see Fig.~S5 in Supplemental
Material~\cite{supplmat}), which is in reasonable agreement with the value of $T_c=390\,\text{K}$ obtained in Ref.~\cite{ershadrad2022unusual}.

\begin{figure*}[t]
	\centering
	\includegraphics[width=1.0\linewidth]{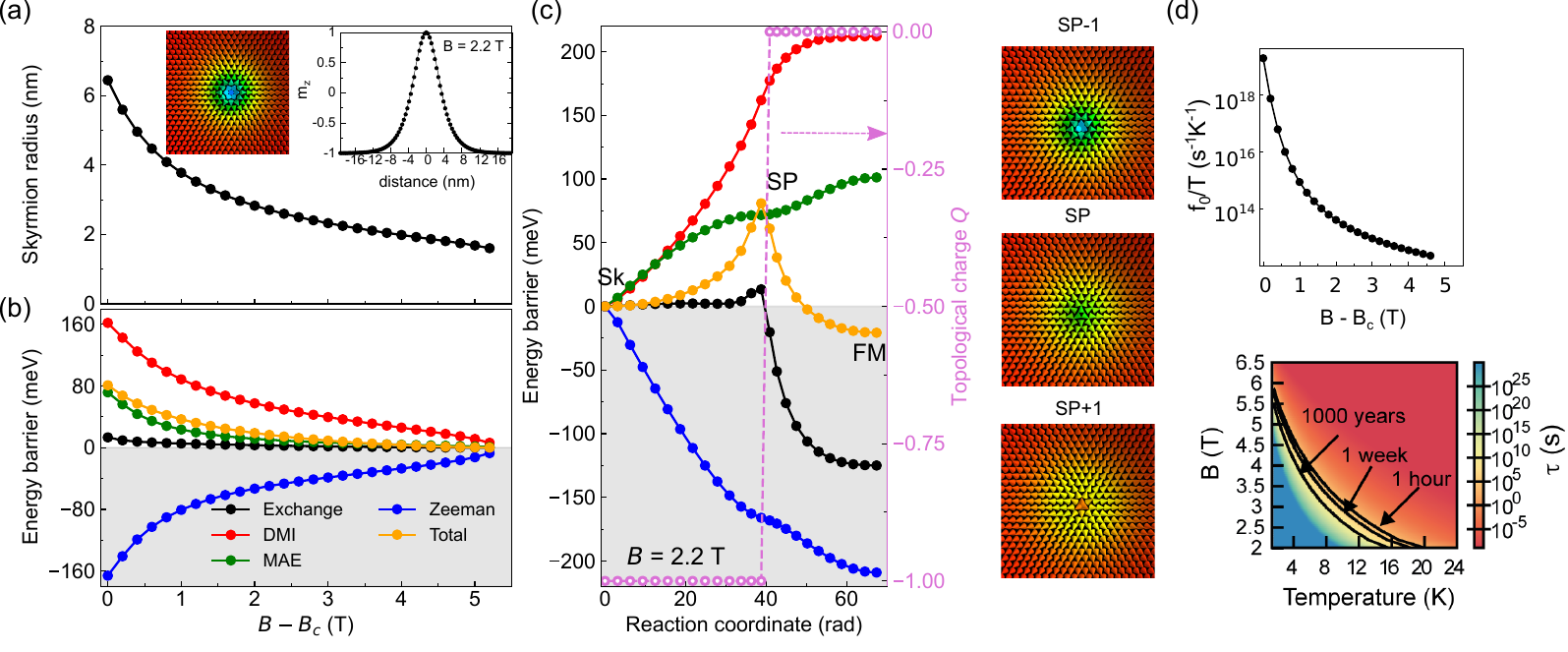}
	\caption{\label{gneb_lifetime} Skyrmion radius and energy barrier for skyrmion collapse in the Fe$_5$GeTe$_2$ monolayer evaluated using magnetic interaction parameters from DFT. (a) Skyrmion radius as a function of $B-B_c$ where $B_c$ is the critical field for stabilizing metastable magnetic skyrmions. The calculated value is $B_c = 2.2$~T for Fe$_5$GeTe$_2$. The skyrmion profile and spin texture at $B = 2.2$~T are shown as insets. (b) Total energy barriers and their decomposition (see legend) of isolated skyrmions versus $B-B_c$. (c) Energy contributions from the different interactions (the same legend as in panel (b), left axis) are shown versus the reaction coordinate along the minimum energy path from the initial (skyrmion) state to the final (FM) state through the saddle point (SP). The energies are summed over all atoms of the simulation box and are given relative to the energies of the initial isolated skyrmion state. The topological charge (open circles,  right axis) is plotted versus the reaction coordinate. (c) Corresponding spin structures before (SP-1), after (SP+1), and at the saddle point (SP) are shown. Note that the skyrmion collapse occurs via radially symmetrical shrinking. (d) Calculated attempt frequencies $f_0$ on a logarithmic scale with respect to $B-B_c$ (top panel). Skyrmion lifetimes of Fe$_5$GeTe$_2$ obtained in harmonic transition state theory based on the spin model with DFT parameters as a function of magnetic field and temperature (bottom panel).}
\end{figure*}

To check the possibility of stabilizing nanoscale magnetic skyrmions in Fe$_5$GeTe$_2$, we performed atomistic spin simulations using the spin model described by Eq.~(\ref{model}) 
with the full set of DFT parameters. We apply
atomistic spin-dynamics 
via the Landau–Lifshitz equation to obtain isolated magnetic skyrmions. 
Minimum energy paths between the initial skyrmion and final FM state
were calculated using the geodesic nudged elastic band (GNEB) method~\cite{bessarab2015method}.
The energy barriers stabilizing skyrmions against collapse were obtained from the saddle point along this path. Finally, we used harmonic transition-state theory to quantify skyrmion stability by calculating their lifetime~\cite{bessarab2018lifetime,Haldar2018,Malottki2019,muckel2021experimental} (see Supplemental Material for computational details~\cite{supplmat}).

\begin{figure*}[tp]
	\centering
	\includegraphics[width=0.80\linewidth]{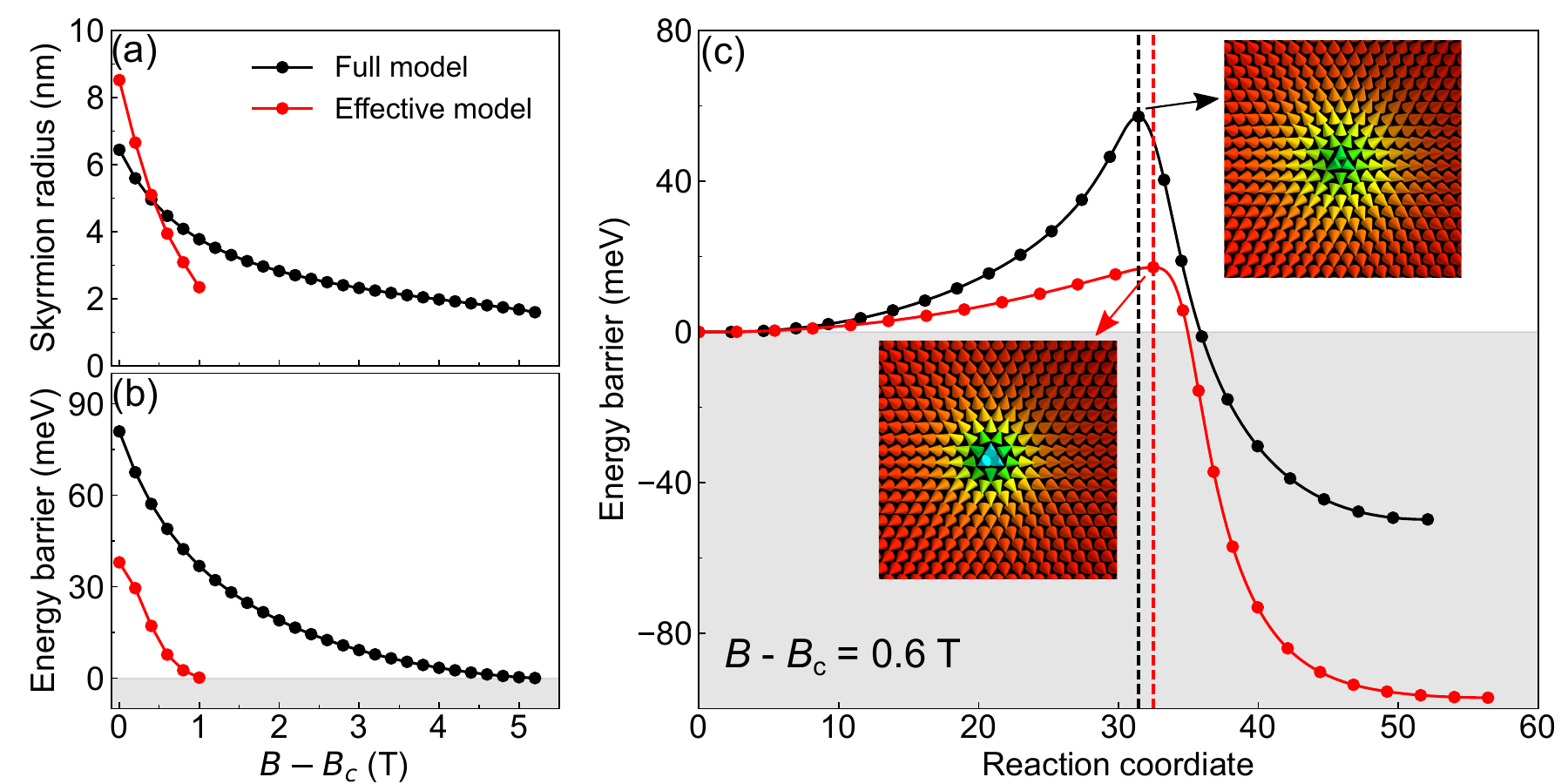}
	\caption{\label{effectivemodel} Full model versus effective model comparison. (a) Skyrmion radius vs.~magnetic field. (b) Skyrmion energy barriers with respect to B. The estimated critical magnetic fields $B_c$ are 2.2 T and 1.4 T for the full model and effective model, respectively. (c) Minimum energy paths of skyrmion collapse: Energies of the spin configurations during a skyrmion collapse are shown over the reaction coordinate corresponding to the progress of the collapse at $B - B_{c} = 0.6$ T. Our simulations were conducted using both the effective exchange model (in red) and considering all magnetic interactions from the DFT calculation (in black).}\end{figure*}

We create isolated skyrmions in the field-polarized background with an out-of-plane magnetization direction due to an applied magnetic field and fully relax these spin structures by solving the damped Landau–Lifshitz equation self-consistently. The in-plane FM state (denoted as FM$_{\parallel}$) exhibits slightly lower energy compared to the out-of-plane FM state, owing to the in-plane MAE of Fe$_5$GeTe$_2$ (see Fig.~S6 in Supplemental Material ~\cite{supplmat}). 
At zero magnetic field, we do not observe the emergence of skyrmions. Instead, we observe labyrinth domains with chiral Néel domain walls, which hold the lowest energy state. As we gradually increase the magnetic field $B$ perpendicular to the monolayer, the labyrinthine domain shrinks and eventually vanishes, giving rise to isolated magnetic skyrmions at a critical field of $B_c$ = 2.2 T. Note, that for fields above $B=2K/M \approx 1.8$~T applied perpendicular to the film the Zeeman energy exceeds the MAE \footnote{Note, that the dipole-dipole interaction also prefers an in-plane magnetization 
direction as the MAE. For monolayer Fe$_5$GeTe$_2$ its energy contribution amounts to about 60\% of the MAE. If we include the dipolar energy, the critical magnetic field required to obtain the field-polarized phase with an out-of-plane magnetization increases to about 3.4~T.}.

Our atomistic spin simulations predict Néel-type magnetic skyrmions stabilized in the FM background (see inset of Fig.~\ref{gneb_lifetime}(a) for skyrmion profile and spin texture at $B=2.2$~T). As expected, the skyrmion size decreases in Fe$_5$GeTe$_2$ with increasing magnetic field (Fig.~\ref{gneb_lifetime}(a)).
Interestingly, nanoscale skyrmions occur with a radius 
below 6.4~nm at $B > 2.2$~T, and isolated skyrmions can be obtained.
Note that the skyrmion radius is estimated using the Bocdanov-$\Theta$ profile \cite{bocdanov1994properties}
within a wide range of $B = 2.2 \sim 7.6$~T. 
At $B> 7.6$~T, skyrmions collapse into the FM state.

To provide insight into skyrmion stability in monolayer Fe$_5$GeTe$_2$, we show in Fig.~\ref{gneb_lifetime}(b) the calculated energy barriers protecting skyrmions from
collapsing into the FM state with respect to $B-B_c$.
Remarkably, we obtain an energy barrier 
of more
than 80 meV 
at the critical field $B_c=2.2$~T. 
This value is comparable to state-of-the-art ultra-thin films that serve as prototype systems to host nanoscale skyrmions \cite{Malottki2017,Haldar2018,paul2020role}. 
Note, that previous work 
on skyrmion stability in 2D magnets \cite{Dongzhe2022_fgt}
reported smaller values
than those found here for the Fe$_5$GeTe$_2$ monolayer. 

From the energy decomposition of the barrier, we 
conclude that the DMI and MAE are mainly responsible for the skyrmion stability
(Fig.~\ref{gneb_lifetime}(b)). We also note that the energy barrier can be 
enhanced if we increase the in-plane MAE, a quantity that can be easily tuned in experiments by doping or temperature \cite{tan2018hard,wang2020modifications,Park2020_nl}. However, if the in-plane MAE is increased larger
magnetic fields are needed to obtain the field-polarized phase in which skyrmions can be stabilized.

To obtain information about transition mechanisms from the skyrmion state to the FM state via the saddle point (SP), which determines the barrier,
we show in Fig.~\ref{gneb_lifetime}(c) the decomposition of the energy along the minimum energy path for skyrmion collapse at $B = 2.2$~T. The topological charge, calculated by $Q=\int \V{m} \cdot (\frac{\partial \V{m}}{\partial x} \times \frac{\partial \V{m}}{\partial y})dxdy$, changes from $-1$ to $0$ at the SP. The skyrmion is annihilated via the radial symmetric collapse mechanism in which the skyrmion shrinks symmetrically to SP and then collapses into the FM state~\cite{muckel2021experimental}. Again, it is clear that the DMI and MAE prefer the skyrmion (Sk) state and decrease the total barrier, while the Zeeman term strongly favors the FM state. Due to frustration, the Heisenberg exchange energy gives a small positive annihilation barrier.

The stability of metastable magnetic skyrmions can be quantified by their mean lifetime, $\tau$, which is given by the Arrhenius law $\tau = f_{0}^{-1} \exp(\frac{\Delta E}{k_{\V{B}} T})$, where $\Delta E$, $f_{0}$, and $T$ are energy barrier, attempt frequency, and temperature, respectively. The calculated $f_{0}$ within harmonic transition state theory~\cite{bessarab2018lifetime} is shown in 
Fig.~\ref{gneb_lifetime}(d). As expected, $f_{0}$ depends strongly on the magnetic field. This effect is similar to that observed in ultrathin transition-metal films, which can be traced back to a change of entropy with skyrmion radius and profile~\cite{Malottki2019,varentcova2020toward}. From the temperature and field dependence of the skyrmion lifetime (Fig.~\ref{gneb_lifetime}(d)), we predict that isolated skyrmions in the Fe$_5$GeTe$_2$ are stable up to hours at a temperature at about 20~K and $B=2.2$~T. Therefore, these nanoscale skyrmions can be probed by experiments using current state-of-the-art techniques e.g., spin-polarized scanning tunneling microscopy or Lorentz transmission electron microscopy.

We plot in Fig.~\ref{effectivemodel} the comparison of atomistic spin simulation results from parameters with the full and effective models. The critical magnetic field is found to be about 
1.4~T in the case of the effective model, which is 0.8~T smaller than the one for the full model. In both models, we have sub-10 nm radius skyrmions 
(Fig.~\ref{effectivemodel}(a)) and energy barriers of a few 10 meV 
(Fig.~\ref{effectivemodel}(b)). From the GNEB calculation at 
$B - B_c = 0.6$~T, the energy barriers for the full and effective models are rather different. We find more than 3 times larger energy barrier with the full model, indicating the importance of including full magnetic parameters for the Fe$_5$GeTe$_2$ monolayer.

In summary, we propose, based on first-principles calculations and atomistic spin simulations,
that monolayer Fe$_5$GeTe$_2$ is a compelling 2D vdW magnet with skyrmionic physics. Due to the large DMI together with moderate in-plane MAE, Fe$_5$GeTe$_2$ monolayer can exhibit metastable nanoscale (sub-10~nm) skyrmions in an out-of-plane magnetic field above 2.2~T. We also highlight the importance of including beyond nearest neighbor DMI terms in the atomistic spin model. The energy barriers protecting skyrmions against collapse are up to 80~meV, which are comparable to those of state-of-the-art transition-metal ultrathin films. Using harmonic transition-state theory, we predict that nanoscale skyrmions are stable in monolayer Fe$_5$GeTe$_2$ with lifetimes of hours up to 20~K. 
	
	\section*{Acknowledgments}
	
This study has been supported through the ANR Grant No. ANR-22-CE24-0019. This study has been (partially) supported through the grant NanoX no.~ANR-17-EURE-0009 in the framework of the ``Programme des Investissements d’Avenir". S.~Ha and S.~He.~gratefully acknowledge financial support from the Deutsche Forschungsgemeinschaft (DFG, German Research Foundation) through SPP2137 ``Skyrmionics" (project no.~462602351). {H.~S.~would like to acknowledge financial support from the Icelandic Research Fund (Grant No. 239435). This work was performed using HPC resources from CALMIP (Grant No. 2021/2023-[P21023]), the North German Supercomputing Alliance (HLRN), and the Kiel University Computing Centre. 
	
	
	\bibliography{References}
	
\end{document}